\documentclass{ws-ijmpa}

\begin{document}


%
\catchline{}{}{}{}{}
%

\title{INSIGHTS INTO THE $\pi^{-}p \to \eta n$ REACTION MECHANISM}

\author{\footnotesize J. DURAND$^{\triangle}$
B. JULIA-DIAZ$^{\circ , \bullet}$, 
T.-S. H. LEE$^{\times , \bullet}$,
B. SAGHAI$^{\triangle}$,
and
T. SATO$^{{\triangleleft , \bullet}}$
}
\address{
$^{\triangle}$Institut de Recherche sur les lois Fondamentales 
de l'Univers, DSM/IRFU, CEA/Saclay, F-91191 Gif-sur-Yvette, France\\
$^{\circ}$Departament d'Estructura i Constituents de la Mat\`{e}ria and Institut 
de Ci\`encies del Cosmos, Universitat de Barcelona, E-08028 Barcelona, Spain\\
$^{\bullet}$Excited Baryon Analysis Center (EBAC), Thomas Jefferson National
Accelerator Facility, Newport News, Virginia 22901, USA\\
$^{\times}$Physics Division, Argonne National Laboratory, 
Argonne, Illinois 60439, USA\\
$^{\triangleleft}$Department of Physics, Osaka University, Toyonaka, 
Osaka 560-0043, Japan\\
}

\maketitle


\begin{abstract}
A dynamical coupled-channels formalism is used to investigate the $\eta-$meson production 
mechanism on the proton induced by pions, in the total center-of-mass energy region
from threshold up to 2 GeV. 
We show how and why studying exclusively total cross section data might turn out to be
misleading in pinning down the reaction mechanism.

\end{abstract}


\section{Introduction}	

Recent extensive phenomenological studies\cite{Durand:2008es} of the process 
$\pi^{-}p \to \eta n$ are motivated not only for its interest {\it per se}, 
but also by the ongoing development of sophisticated coupled-channels formalisms 
in order to determine the properties of baryon resonances\cite{Matsuyama:2006rp}.

In this contribution we concentrate on the double bump structure observed in
the total cross section ($\sigma_{tot}$) of the $\pi^{-}p \to \eta n$ reaction. 
The first maximum
is unambiguously generated by the $S_{11}(1535)$ resonance, while the origin of the
second one is still not well established. Here, we give a very brief account
of published findings. All those models include nonresonant terms, and the
resonances
$S_{11}(1535)$ and $S_{11}(1650)$ (hereafter called core terms), but differ 
in additional resonances and/or
the extent of coupled-channels content. With respect to this latter point, they all embody
$\pi N$ and $\eta N$, and in some cases $\pi \pi N$ {\it via} $\pi \Delta$, 
$\sigma N$, and $\rho N$ intermediate-states. Within an early K-matrix approach,
Sauermann {\it et al.}\cite{Sauermann:1994pu}, using the core terms, find no second bump,
which appears by adding the $P_{13}(1720)$.
Gridnev and Kozlenko\cite{Gridnev:1999sz} work based also on the K-matrix, produces
a double bump structure in the S-wave only, but the minimum turns out to be roughly 
two orders of magnitude too low. Penner and Mosel\cite{Penner:2002ma} introduce a more
elaborated K-matrix coupled-channels with the above mentioned five intermediate-states 
plus $\omega N$, $K \Lambda$, and $K \Sigma$. They attribute the second maximum to the
$P_{11}$-wave and get a good agreement with the data by including also the
$P_{13}$- and $D_{13}$-wave resonances. A direct-channel constituent quark
model\cite{Zhong:2007fx} finds also $P_{11}$-wave crucial with respect to the second
bump.
Finally, Gasparyan {\it et al.}\cite{Gasparyan:2003fp}, in a more comprehensive
version of the J\"{u}lich meson-exchange model, obtain a good agreement with the data
{\it via} the core terms plus the $P_{13}(1720)$, with small contribution from the
$D_{13}(1520)$, but their angular distributions for the  $d\sigma / d\Omega$ 
deviate (significantly) from the data above $W \approx$ 1.65 GeV.

In Sec. 2 we outline our approach and in Sec. 3 our findings are presented, 
showing how the interplay of various resonances might lead to different conclusions 
through $d\sigma / d\Omega$ or $\sigma_{tot}$.
%
%
\section{Formalism and model}
A dynamical coupled-channels formalism\cite{Matsuyama:2006rp}, proven to be successful 
in studying the $\pi N \to \pi N$ reactions\cite{JuliaDiaz:2007kz}, is used to 
investigate\cite{Durand:2008es} the $\eta-$meson production on the proton induced 
by pions. 
The coupled-channels equations are derived from standard projection operator techniques.
The nonresonant interactions are deduced from a unitary transformation method, applied 
on a set of phenomenological Lagrangians\cite{Matsuyama:2006rp}.
This approach includes intermediate $\pi N$, $\eta N$, $\pi \Delta$, $\sigma N$, and $\rho N$ 
channels and all three and four star resonances with $ M \le$ 2 GeV, namely,
$S_{11}(1535)$, $S_{11}(1650)$, $P_{11}(1440)$, $P_{11}(1710)$, $P_{13}(1720)$,
$D_{13}(1520)$, $D_{13}(1700)$, $D_{15}(1675)$, and $F_{15}(1680)$.
The model $B$ reported in Ref.~[1] is used in the present work, and
hereafter called the full model.
That model is obtained by fitting {\it exclusively} the $d\sigma / d\Omega$ data 
for the reaction $\pi^- p \to \eta n$ ($W \lesssim$ 2 GeV), leading to a reduced $\chi^2$ = 1.96.
Consequently, the $\sigma_{tot}$ results reported in the next section are predictions
from that model.
%
%
\section{Results and discussion}
\begin{figure}[t]
\begin{center}
\parbox[c]{0.7\textwidth}{
\centering
\includegraphics[width=0.7\textwidth]{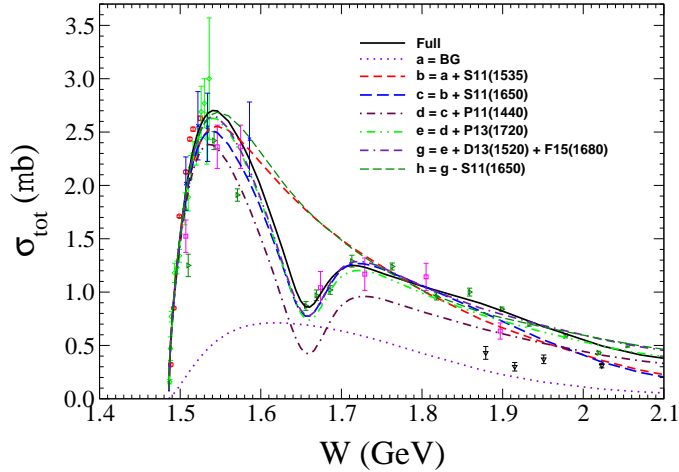}
}
\caption{Total cross section for the process $\pi^- p \to \eta n$ as a function
of total c.m. energy. 
Curves (a) to (g) are obtained for background, and then adding one after another
resonances $S_{11}(1535)$, $S_{11}(1650)$, $P_{11}(1440)$, 
$P_{13}(1720)$, and
$D_{13}(1520)$ plus $F_{15}(1680)$.
Curve (h) contains all those resonances except the $S_{11}(1650)$. 
The full model  embodies all those terms plus 
the $D_{13}(1700)$ and $D_{15}(1675)$. 
Data\protect\cite{Data}  are from
Deinet~{\it et al.} (crosses), 
Brown~{\it et al.} (right triangles),
Crouch {\it et al.} (down triangles),
Debenham~{\it et al.} (up triangles), 
Morrison (diamonds)
and Prakhov~{\it et al.} (empty circles).
}
\label{F1}
\end{center}
\end{figure}

Total cross section as a function of total c.m. energy is depicted in
Fig.~\ref{F1}. The full model describes satisfactorily the data. The main feature
of the data is two bumps at around 1.560 and 1.710 GeV, with a minimum at roughly
1.660 GeV. 
To get deeper insights into the structure of the $\sigma_{tot}$, we start with results
from the background terms and show contributions from the most significant 
resonances\cite{Durand:2008es} introducing them one after another.

The background terms produce a smoothly varying behavior, the value of which 
becomes sizable close to the observed minimum. Adding the $S_{11}(1535)$ 
to this latter
gives the essential features of the $\sigma_{tot}$, especially the position and the
size of the first peak, but overestimates the data in the range 
1.6 $\lesssim W \lesssim$ 1.7 GeV. 
By adding on top of the previous terms the 
$S_{11}(1650)$, the full model's results are almost recovered. Accordingly, the
minimum emerges from destructive interference between the two 
$S_{11}$-resonances. 
Hence, the second maximum appears just because of vanishing
contribution from the second $S_{11}$-resonance for $W \gtrsim$ 1.7 GeV, 
its magnitude being hence
produced by the first $S_{11}$-resonance. Introducing additional resonances,
the shape is not altered. Actually, the decrease of the $\sigma_{tot}$
because of the $P_{11}(1440)$ is compensated by the $P_{13}(1720)$, while
the $D_{13}(1520)$ and $F_{15}(1680)$ introduce small contributions. The effects
of the $D_{13}(1700)$ and $D_{15}(1675)$, found negligible\cite{Durand:2008es}, 
are not depicted.
Finally, we show the results with all above terms except the $S_{11}(1650)$,
endorsing the observation that the structure is due to the interference
between the two $S_{11}$-resonances.

At the present stage and based
on the $\sigma_{tot}$, one could conclude that the reaction mechanism
involves merely background terms and the lowest lying $S_{11}$-resonances.
To avoid such a misleading conclusion, we move to the $d\sigma / d\Omega$. 
Here we single out three energies corresponding
to the positions of the first and the second maximums, as well as to the $W$, where
the $\sigma_{tot}$ comes down to its value of the minimum. There are no data at those
energies. However, our model has been successfully compared\cite{Durand:2008es} 
to all the relevant $d\sigma / d\Omega$ data.
%

\begin{figure}[t]
\begin{center}
\parbox[c]{0.97\textwidth}{
\centering
\includegraphics[width=0.97\textwidth]{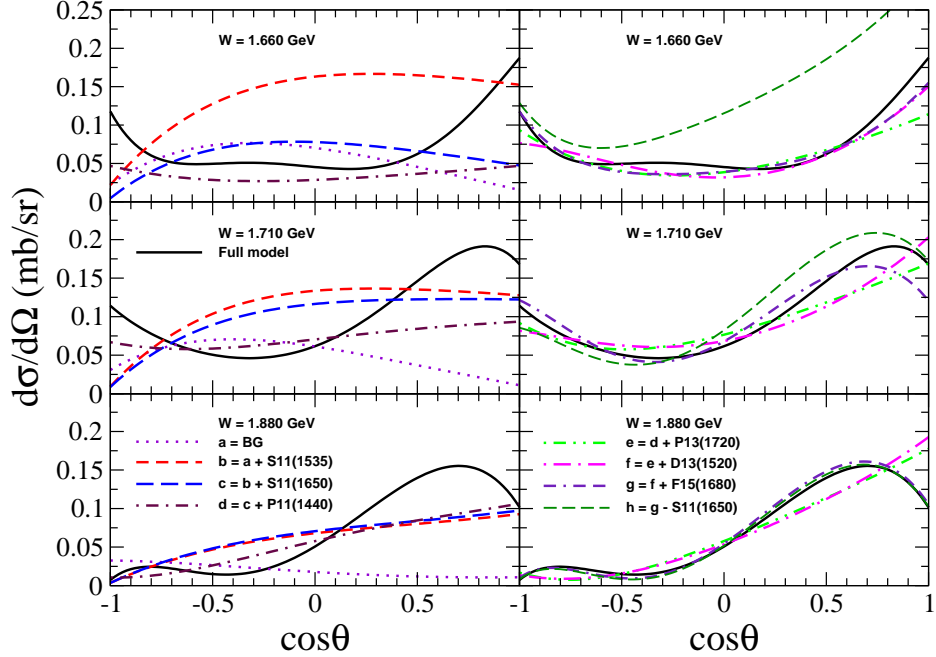}
}
\vspace*{8pt}
\caption{Differential cross-section angular distribution for the process 
$\pi^- p \to \eta n$ at three energies: $W$=1.660, 1.710, 1.880 GeV.
Curves notations are the same as in Fig.~\ref{F1}. 
}
\label{F2}
\end{center}
\end{figure}

Figure~\ref{F2} shows our results for the $d\sigma / d\Omega$ at the above three
energies. The background behaves smoothly and decreases with increasing energy.
The $S_{11}(1535)$ brings in the dominant contribution, while the $S_{11}(1650)$
has a destructive effect, which is very significant at the lowest energy depicted,
explaining the minimum found in the $\sigma_{tot}$. It is instructive to notice 
that, contrary to the $\sigma_{tot}$,
the sum of those three terms (curve c) is far away from the full model and gives a wrong curvature. The $P_{11}(1440)$ introduces also 
a destructive contribution at the two lowest energies, and, more importantly, 
reverses the curvature of the $d \sigma / d \Omega$. While the $P_{13}(1720)$ amplifies
the latter behavior, the $D_{13}(1520)$ has a small effect. Finally, the correct shape
is induced by the $F_{15}(1680)$. The last curve (h), contains all those resonances
except the $S_{11}(1650)$. The corresponding curve shows very significant
deviations from the full model at the lowest energy. That effect is suppressed at
the next energy and vanishes at the highest one.

In summary, using a dynamical coupled-channels approach embodying
all known three and four star resonances and reproducing satisfactorily differential and
total cross section data for the process $\pi^{-}p \to \eta n$, 
we have shown that (i) the structure of the $\sigma_{tot}$
is dictated by the $S_{11}(1530)$ and $S_{11}(1650)$ interference,
(ii) those $S$-wave resonances are by far insufficient to reproduce
the differential cross section data.
 
\end{document}